\documentclass[useAMS,usenatbib]{mn2e}

\title[Dust Penetrated Arm Classes]{Dust Penetrated Arm Classes: Insights 
from rising and falling rotation curves}

\author[M.~S. Seigar et al.]{M.~S. Seigar$^{1, 2}$\thanks{E-mail: 
mseigar@uci.edu (MSS)}, D.~L. Block$^{3}$, I. Puerari$^4$, N.~E. 
Chorney$^{2, 5}$ and P.~A. James$^{6}$\\
$^{1}$Department of Physics \& Astronomy, University of California Irvine,
4129 Frederick Reines Hall, Irvine, CA 92697-4575, USA\\
$^{2}$Joint Astronomy Centre, 660 N. A'ohoku Place, Hilo, HI 96720, USA\\
$^{3}$School of Computational and Applied Mathematics, University of the 
Witwatersrand, P.~O. Box 60, Wits, Gauteng 2050, South Africa\\
$^{4}$Instituto Nacional de Astrof\'isica, Optica y Electr\'onica, Apdo. 
Postal 51 y 216, 72000 Puebla, Pue., Mexico\\
$^{5}$Department of Physics \& Astronomy, University of Victoria, PO Box 3055 
STN CSC, Victoria BC, V8W 3P6, Canada\\
$^{6}$Astrophysics Research Institute, Liverpool John Moores University, 
Twelve Quays House, Egerton Wharf, Birkenhead CH41 1LD}

\begin{document}

\date{In original form 2004 October 3}

\pagerange{\pageref{firstpage}--\pageref{lastpage}} \pubyear{2004}

\maketitle

\label{firstpage}

\begin{abstract}
In the last decade, near-infrared imaging has highlighted the
decoupling of gaseous and old stellar disks: the morphologies of 
optical 
(Population I) tracers compared to the old stellar disk morphology, 
can be radically
different. Galaxies which appear multi-armed and even flocculent in 
the optical may show significant Grand-Design spirals in the near-infrared. 
Furthermore, the optically determined Hubble 
classification
scheme does not provide a sound way of classifying dust-penetrated
stellar disks: spiral arm pitch angles (when
measured in the near-infrared) do not correlate with Hubble type. The 
dust--penetrated classification scheme of Block \& Puerari  provides
an alternative  classification based on near-infrared morphology, and which is 
thus more closely linked to the dominant stellar mass component.
Here we present
near--infrared K band images of 14 galaxies, on which we have
 performed a  Fourier 
analysis of the spiral structure in order to determine
their near-infrared pitch angles and dust--penetrated arm classes. 
We have also used the rotation curve data of Mathewson et al 
to  calculate the rates of shear in the stellar disks of these galaxies.
We find a correlation between near-infrared pitch angle and rate of shear: 
galaxies
with wide open arms (the $\gamma$ class) are found to have rising
rotation curves, while those with
falling rotation curves belong to the tightly wound $\alpha$ bin.
The major determinant of near-infrared spiral arm pitch angle is the 
distribution of matter within the galaxy concerned. The correlation 
reported in this study provides the physical basis underpinning 
spiral arm classes in the
dust-penetrated regime and underscores earlier
spectroscopic findings by Burstein and Rubin that Hubble type and mass
distributions are unrelated.
\end{abstract}


\section{Introduction}

The classification of galaxies by Hubble type (Hubble 1926),
is traditionally inferred in the
optical regime, where dust extinction
still has a large affect, and where the light is dominated by the young
Population I stars. 
Infrared arrays offer unique opportunities for
decoupling the Population I and Population II disk morphologies, because
in the K-band (2.2 $\mu$m), dust extinction is minimal, and the light
is dominated by old and intermediate age stars (Rhoads 1998; Worthey 1994).
 The extinction at this
wavelength is only 10\% of that in the V-band (Martin \& Whittet 1990).

Optically determined Hubble
type is not correlated with dust-penetrated Population II morphology, as
confirmed by the near-infrared studies of de Jong (1996) and Seigar \& James
(1998a, b). 
Even in the optical regime, the correlation between quantitative pitch angle
and Hubble type is
weak (see Kennicutt 1981), even though tightness of the spiral arm pattern,
as
judged by eye, is one of the defining Hubble morphological criteria.
Furthermore, it has been shown that near-infrared morphologies of spiral
galaxies can be significantly different from their optical 
morphologies (Block \&
Wainscoat 1991; Block et al. 1994a; Thornley 1996; Seigar \& James 1998a, b;
Seigar, Chorney \& James 2003). Galaxies with flocculent spiral
structure in the optical may present Grand-Design spiral structure in the
near-infrared (Thornley 1996; Seigar et al. 2003). Such studies
suggest that the
optical morphology gives only incomplete and sometimes deceptive information
about the underlying stellar mass distribution.

Burstein \& Rubin (1985) showed that spiral galaxies can belong to one of 
three form families, based on the shapes of their rotation curves: 
whether rising, falling or flat. Burstein and Rubin differentiated
three principal types of mass distribution and found
that Hubble types Sa and Sb were represented amongst all three types 
in approximately equal numbers. This early spectroscopic 
study was a strong pointer to the decoupling of
gaseous and stellar disks confirmed by near-infrared imaging.

From a dynamical viewpoint, the disk of a spiral galaxy can be separated into
two distinct components: the {\em gas-dominated} Population I disk, and the
{\em star-dominated} Population II disk. The former component contains
features of spiral structure (OB associations, HII regions, and cold
interstellar HI gas), which are naturally fast evolving; dynamically, it is
very active and responsive, because, being characterized by small random
motions (i.e. a `cool disk'), it fuels Jeans instability. In contrast, the
Population II disk, which is dynamically `warmer', 
contains the old stellar population highlighting the underlying
stellar mass distribution (Lin 1971). One might expect, even in the absence of
appreciable optical depths, for the two morphologies to be very different,
since the near-infrared light comes predominantly from intermediate-age 
giant and supergiant stars (Rix \& Rieke 1993; Frogel et al. 1996). 

It is important to stress, that from this dynamical viewpoint, one requires
two classification schemes, a near-infrared classification scheme, such as the
dust-penetrated class (Block \& Puerari 1999; Block et al. 1999) to describe 
the Population II disk, as well as optically determined Hubble type to 
describe the Population I 
disk. A near-infrared classification scheme can never replace an optical one,
and vice--versa, because the current distribution of old stars strongly affects
the distribution of gas in the Population I disk. The dynamic interplay 
between
the two components (via a feedback mechanism) is crucial, and has been
studied extensively (Bertin \& Lin 1996). A central aspect here is the likely
coupling of the Population I disk with that of the Population II disk via a
feedback mechanism (Pfenniger et al. 1996). To derive a coherent 
physical framework for the excitation of spiral structure in galaxies, one 
must consider the co-existence of the two dynamical components.

There is a fundamental limit in predicting what evolved stellar disks might 
look like. The greater the degree of decoupling, the greater is the 
uncertainty. The fact that a spiral might be flocculent in the optical is 
very 
important, but it is equally important to know whether or not there is a 
decoupling with a Grand-Design old stellar disk. No prediction on that issue 
can, a priori, be made (Block, Elmegreen \& Wainscoat 1996).

The theoretical framework to explain the co-existence of completely different
morphologies within the same galaxy when studied optically and in the
near-infrared is described by Bertin \& Lin (1996), following the pioneering
work of Lindblad (1963).
A global mode (Bertin et al. 1989a, b) is composed
of spiral wavetrains propagating radially in opposite
directions, similar to a standing wave. Thus a feedback of wavetrains is
required from the centre. The return of wavetrains back to the corotation
circle is guaranteed by refraction, either by the bulge or because the inner
disk is dynamically warmer. In the stellar disk, such a feedback can be
interrupted by the Inner Lindblad Resonance (ILR), which is a location where
the stars meet the slower rotating density wave crests in resonance with their
epicyclic frequency (Mark 1971; Lynden-Bell \& Kalnajs 1972). In the gaseous
disk, the related resonant absorption is only partial, so that some feedback
is guaranteed. Once the above described wavecycle is set up (in the absence of
a cutoff by an ILR), a self-excited global mode can be generated.

The tightness of the arms in the modal theory comes from the mass distribution
and rate of shear. Galaxies with a more central
mass concentration, i.e. higher overall
densities (including dark matter) and higher rates of shear, 
are predicted to have more tightly wound arms. The models of Fuchs (1991,
2000) result in disks with rigidly rotating spiral modes, wherein bulges act
as inner reflectors of waves or modes induced by the swing amplification
method, thus leading to modal spiral waves, which form as a result of Toomre's
swing amplification (Toomre 1981). Fuchs (2000) adopts a stellar-dynamical
analogue of the Goldreich \& Lynden-Bell (1965) sheet, which describes the
local dynamics of a patch of thin, differentially rotating stellar disk. The
result is that, instead of shearing density waves (as in the unbounded sheet),
spiral modes do appear. It can be shown that it is still the swing 
amplification mechanism, which is responsible for the appearance of the modes
in these half--bounded disks. The models of Fuchs are disks which rigidly
rotating spiral modes, wherein bulges act as inner reflectors of waves induced
by swing amplification (see also Bertin \& Lin 1996).
These models best show how central mass concentration
correlates with spiral arm pitch angle.
If the disk is very light (low $\sigma$ where $\sigma$ is the disk density) 
the mode can be very tight, and one is in the domain of small epicycles
(formally, the stability parameter $Q=c \kappa / \pi G \sigma$ 
being close to unity,
the value of $c$ must also be small, where $c$ is the radial velocity 
dispersion and $\kappa$ is the epicyclic frequency).
If one increases the mass of the disk one finds a trend towards more open
structures, but soon one runs the risk of a disk that is too heavy and a bar
mode results. The trend towards more open spiral structure also follows a
trend towards smaller rates of shear.

In those spirals with more open spiral arms at K but with no sign of a 
prominent bar, Bertin (1996) anticipates that the galaxy should be gas rich. 
Abundant gas can shock, dissipate and make some violently unstable open modes 
(Bertin \& Lin 1996). This is one important reason why `gas content' is 
important in the framework of modal theory for classification as the 
governing 
parameter for the trend from early-type (i.e. `a') to late-type
(i.e. `c') spiral galaxies (Block et al. 1994a).

The redistribution of angular momentum by large-scale spiral torques will be 
stronger for stellar arms which are more open; some authors (eg. Pfenniger et 
al. 1996) have postulated that such a redistribution may lead to rapid 
changes in the disk with more mass being concentrated towards the centre, 
and even modify the properties of the rotation curve. 
This is the concept of secular evolution of a galaxy, from an open to a more 
tightly wound morphology, within one Hubble time.

Many galaxies show the presence of a significant $m$=1 component in the 
near--infrared (often in the form of a lopsidedness of the spiral). The 
linear modal theory predicts that $m$=1 modes should generally be dominated 
by $m$=2 modes when available, since the latter are more efficient in 
transporting angular momentum outwards. However, modes greater than $m$=2 are 
generally suppressed in the stellar disk by an ILR. While the disk mass 
participating in the mode is crucial, the gas--content of the galaxy is 
important: gas--rich spirals can generate modes greater than $m$=2. It had 
earlier been predicted (Block et al. 1994a, Bertin \& Lin 1996) that infrared 
images should show an ubiquity of global one and two armed structures in the 
underlying stellar disk and the dust--penetrated classification scheme does 
confirm such a trend.

In contrast, the dynamics of the cold Population I gaseous disk, 
characterized 
by a different scalelength, velocity dispersion, thickness, and behaviour at 
the relevant Lindblad resonances, explains why spiral galaxies are optically 
so often overwhelmed by higher $m$ modes and other more irregular fast 
evolving features, supported by the cold interstellar gas (Bertin 1991, 1993).

The dust-penetrated classification scheme (Block \& Puerari 1999; Block et al.
1999) is a quantitative way of classifying galaxies 
according to their near-infrared spiral arm pitch angles
(measured from 2 $\mu$m images). 
Details of this classification scheme are elucidated in section 3.

\begin{figure*}
\vspace*{25cm}
\end{figure*}

\begin{figure*}
\vspace*{25cm}
\end{figure*}

\begin{figure*}
\vspace*{8cm}
\caption{Greyscale images of the galaxies for which the FFT analysis was 
performed. The overlaid contours represent the FFT fit to the spiral 
structure.}
\end{figure*}

In this paper, we present near-infared images for 14 spiral galaxies. We
assign dust--penetrated classes to all of them and show that a correlation
exists between the shear rate in stellar disks (as derived from their 
rotation curves) and their near-infared spiral arm pitch angles. The dust
penetrated arm class depends on the near-infared spiral arm pitch angle, and
so, such a correlation would provide a physical basis  underpinning
the dust penetrated arm scheme of Block \& Puerari (1999). Section 2 
describes the observations; section 3 describes our Fourier analysis and how
we calculate pitch angles and assign dust penetrated classes; section 4 
discusses our results and shows how we calculate shear rates in these galaxies
while section 5 summarises our conclusions.

\section{Observations}

We have observed a sample of 14 galaxies in the near-infrared K-band
(2.2 $\mu$m). These objects were taken from the study of Mathewson et al.
(1992), who measured, by means of long-slit optical spectroscopy, H$\alpha$
rotation curves for 965  southern hemisphere spiral
galaxies. Our sample includes galaxies with
different rotation curve types (rising,
falling and flat) and span as wide a range of optical Hubble types 
as possible. They were also chosen to have moderate inclination angles by
restricting the ratio of the minor axis $b$ to the major axis $a$ using the 
range $0.50<b/a<0.66$. This results in values of inclination from 
$i\sim 42^{\circ} - 60^{\circ}$.

The images were observed at the United Kingdom Infrared Telescope (UKIRT) 
using
the UKIRT Fast Track Imager (UFTI) between 1--4 August 2001, 11--12 March
2002 and 23 February 2004. 
The images from the August 2001 run were observed to a depth of 
21.5 K-mag$/$arcsec$^2$ at the $3\sigma$ level, whereas those from the March
2002 and February 2004 runs were observed down to the same isophote, but at 
the $5\sigma$ level. 

The galaxies observed in the August 2001 run were IC 1330, ESO 602-G25, 
ESO 606-G11, NGC 7677, UGC 14 and UGC 210. The galaxies observed in the
March 2002 run were ESO 515-G3, ESO 574-G33, ESO 576-G51, ESO 583-G2, 
ESO 583-G7 NGC 2584 and NGC 2722. NGC 3456 was observed in February 2004.
The sample is essentially the same as that of Seigar et al. (2003) with 
the addition of NGC 3456 (as it was observed at a later date) and the
omission of ESO 543-G12, ESO 555-G8, ESO 576-G12 and UGC 12383, as 
reliable pitch angles could not be measured for these galaxies.

\section{Decomposition and identification of modes}

The 2-D Fast Fourier decomposition of all the near-infrared images in this
study, employed a program developed by I. Puerari (Schr\"oder et al. 1994).
Logarithmic spirals are assumed in the decomposition. 

The amplitude of each Fourier component is given by:

\begin{equation}
A(m,p)=\frac{\Sigma^{I}_{i=1}\Sigma^{J}_{j=1}I_{ij}(\ln{r},\theta)\exp{-(i
(m\theta+p\ln{r}))}}{\Sigma^{I}_{i=1}\Sigma^{J}_{j=1}I_{ij}(\ln{r},\theta)}
\end{equation}
where $r$ and $\theta$ are polar coordinates, $I(\ln{r},\theta)$ is the
intensity at position $(\ln{r},\theta)$, $m$ represents the number of arms
or modes, and $p$ is the variable associated with the pitch angle $P$, defined
by $\tan{P}=-\frac{m}{p_{max}}$. In this paper we measure the K-band pitch 
angle $P_K$ of the m=2 component.

\begin{figure*}
\vspace*{25cm}
\end{figure*}

\begin{figure*}
\vspace*{18cm}
\caption{Relative Strengths of spiral modes in the galaxies. Plots are in
terms of phase versus relative amplitudes.}
\end{figure*}

Our Fourier spectra corroborate earlier observational indications
(Block et al. 1994a, 1999; Block \& Puerari 1999) that there is indeed a
ubiquity of $m$=1 and $m$=2 modes in the near-infrared. Block \& Puerari
(1999) proposed three
principal archetypes for the evolved stellar disk of such galaxies. The first
of these, designated dust-penetrated class $\alpha$, covers the pitch angle
range $4^{\circ}<P<15^{\circ}$, the second, designated $\beta$, covers
$18^{\circ}<P<30^{\circ}$ and the third, designated $\gamma$ covers
$36^{\circ}<P<76^{\circ}$.

Those galaxies where $m$=1 is the dominant mode are designated
L$\alpha$, L$\beta$ and L$\gamma$ according to the dust penetrated pitch 
angle. Galaxies where $m$=2 is the dominant Fourier mode are classified
into classes E$\alpha$, E$\beta$ and E$\gamma$. Higher
order harmonics are classified as H3 (for $m$=3) and H4 (for $m$=4).

The range of radii over which the Fourier fits were applied are selected to
exclude the bulge or bar (where there is no information about the arms) and
extend to the outer limits of the arms in our images. Pitch angles are then
determined from peaks in the Fourier spectra, as this is the most powerful
method to find periodicity in a distribution (Consid\`ere \& Athanassoula
1998; Garcia-Gomez \& Athanassoula 1993). The radial range over which the
Fourier analysis was performed was chosen by eye and is probably the dominant
source of error in the calculation of pitch angles. As a result three
different radial ranges (listed in Table 1)
were chosen for each galaxy, and a mean pitch angle
and standard error calculated for every object.

It should be noted that bars (often equipped with rather tight spiral arms
outside the bar) form in heavy disks as a natural gravitational instability
(Hohl 1971). However, unbarred spiral structure should be supported
by dynamically light disks. The transition from light to heavy disks and
the affect of this upon bar streangth is described by Bertin (1991) and
Bertin \& Lin (1996).

We are fully aware that the spiral arms in barred galaxies may depart
from a logarithmic shape. The arms may break at a large angle to the
bar and then wind back to the other side, as in a `pseudoring'. Outer
pseudorings (e.g. NGC 3504) present additional complications, but we
do wish to stress that our pitch angles always correspond to arm
morphology outside of the inner bar/bulge or inner pseudoring
region. The robustness of our method and the choice of radial range
may be seen in the ringed SB(r) spiral NGC 5921, where our contours
trace out two grand design stellar arms exterior to the bar and the
inner ring (see Figure 3 in Block et. al. 2001).

The images were firstly deprojected to face-on. Mean uncertainties of position
angle and inclination as a function of inclination were discussed by
Consid\`ere \& Athanassoula (1988). For a galaxy with low inclination,
there are clearly greater uncertainties in assigning both a position angle
and an accurate inclination. These uncertainties are discussed by Block et al.
(1999), who took a galaxy with low inclination ($< 30^{\circ}$)
and one with high inclination ($> 60^{\circ}$)
and varied the inclination angle used in the correction to face-on. They found
that for the galaxy with low inclination, that the measured pitch angle
remained the same. However, the measured pitch angle for the galaxy
with high inclination varied by $\pm$10\%. Since we have only one object with
an inclination angle $> 60^{\circ}$ in our sample (ESO 602 G25) and this
object has an error $>$10\% on the pitch angle anyway, we assume that
inclination affects are not a dominant source of error for these objects.
It should also be noted that our deprojection assumes that spiral galaxy disks are
intrinsically circular in nature. 

Figure 1 shows the images of the spiral
galaxies observed for this project, overlaid with contours representing the
results of an inverse Fourier transform analysis of the $m$=2 component. 
After having deprojected
the K band images and identifying the dominant modes, we calculated the 
inverse Fourier transform, as follows:

We define the variable $u=\ln{r}$. Then
\begin{equation}
S(u,\theta ) = \Sigma_{m}S_{m}(u)\exp{(im\theta)}
\end{equation}
where
\begin{equation}
S_{m}(u)=\frac{D}{\exp{(2u)}4\pi^{2}}\int_{-\infty}^{+\infty}G_{m}(p)A(p,m)
\exp{(ipu)}dp
\end{equation}
and
\begin{equation}
D=\Sigma_{i=1}^{I}\Sigma_{j=1}^{J}I_{ij}(u,\theta)
\end{equation}
$G_{m}(p)$ is a high frequency filter used by Puerari \& Dottori (1992). For 
the spiral with $\tan{P}=-\frac{m}{p_{max}^{m}}$ it has the form:
\begin{equation}
G_{m}(p)=\exp{\left[-\frac{1}{2}\left(\frac{p-p^{m}_{max}}{25}\right)^{2}
\right]}
\end{equation}
where $p_{max}^{m}$ is the value of $p$ for which the amplitude of the Fourier
coefficients for a given $m$ is maximum. This filter is also used to smooth
the $A(p,m)$ spectra at the interval ends (Puerari \& Dottori 1992). The
contour overlays of the inverse Fourier transform in Figure 1 indicate the
excellent fit of our m=2 spiral modes to the spiral structure of the galaxies
observed. The dust-penetrated arm classes and
pitch angles are listed in Table 1.

\section{Rotation Curve Data}

The 14 galaxies observed here all have H$\alpha$ rotation curve data measured
by Mathewson et al. (1992). The rotation curves are presented in Figure 
3. It should be noted that the rotation curves presented here
are symmetrized rotation curves, obtained by averaging the measured values
of rotation relative to the two opposite sides along the major axis.

\begin{figure*}
\vspace*{25cm}
\end{figure*}

\begin{figure*}
\vspace*{5.5cm}
\label{rotation}
\caption{Rotation curves from Mathewson et al. (1992).}
\end{figure*}

As can be seen the rotation curves are of good quality. The errors
on the data points in the rotation curves are a combination of the
error intrinsic to the spectroscopic measurement, which Mathewson
et al. (1992) quote as being $<10$ km s$^{-1}$, and the error
associated with the procedure of folding the two sides of the 
galaxy, which is also typically $<10$ km s$^{-1}$. These rotation
curves have been used to estimate the shear rates in these galaxies
and more detail is given in the Discussion section of this paper.

The work presented by Block et al.
(1999) consisted of only 4 galaxies. Here we present a
spectroscopic/near-infrared imaging analysis of a further 14 galaxies.
The  rates of shear are derived from their
rotation curves as follows
\begin{equation}
\label{shear}
\frac{A}{\omega}=\frac{1}{2}\left(1-\frac{R}{V}\frac{dV}{dR}\right)
\end{equation}
where $A$ is the first Oort Constant, $\omega$ is the angular velocity,
and $V$ is the velocity measured at radius $R$. The value $A/\omega$ gives the
shear rate. 

Using equation \ref{shear}, we have calculated the shear rates for these 
galaxies, over the same radial ranges for which the Fourier analysis was
performed and pitch angles calculated (the radial ranges are listed in
Table 1). We selected several different radial
ranges, just as in the Fourier analysis, and present mean shear rates and
standard errors. The dominant source of error on the shear rate is the
spectroscopic errors given in the rotation curves (i.e. a combination of 
the intrinsic spectroscopic error and the error associated with 
folding the two sides of the galaxy). Typically speaking,
this is $<10$\%. These are listed in Table 1. In order to calculate the
shear rate, the mean value of $\frac{dV}{dR}$ measured in km s$^{-1}$
arcsec$^{-1}$ is calculated by fitting a line of constant gradient to 
the outer part of the rotation curve (i.e. past the radius of turnover).
Average shear rates were then calculated for the rotation curve, over the
radial ranges listed in Table 1. The shear rates listed in Table 1 are
means of the three shear rates measured over the three radial ranges.

\begin{table*}
 \centering
  \caption{Results from the Fourier analysis and rotation curve analysis of 
14 spiral galaxies. Column 1 shows the name of the galaxy; Column 2 shows the 
derived
dust penetrated class; Column 3 shows the optically determined Hubble type; Column 4 shows the 
pitch
angle for the m=2 component of the K-band spiral arms. The error associated with the pitch
angle is related to the radial range over which the FFT is performed. The error is a standard
error derived from the different pitch angles that are calculated when different radial ranges
are used in the FFT analysis. The signal-to-noise ratio is probably the source of the
error shown here.
Column 5 shows the derived shear rate. The error associated with the shear rate is associated
with the intrinsic spectroscopic error and the error associated with the procedure of folding
the two sides of the galaxy.
Column 6 shows the three radial ranges over which the pitch angle and shear rate were determined; 
Column 7 shows the position angle of the major-axis and column 8 shows the 
inclination angle to the plane of the sky.}
  \begin{tabular}{llllllll}
  \hline
  Galaxy Name& Dust--            & Hubble      & $P_K$        & $A/\omega$      & Radial ranges                   & Position              & Inclination   \\
             & penetrated        & Type        &              &                 & in arcsec                       & angle                                 \\
             & class             &             &              &                 & \\
 \hline
ESO 515 G3      & L$\gamma$       & SB(rs)c     & 47.8$\pm$0.7 & 0.27$\pm$0.01  & 13.2-30.9; 12.3-31.8; 14.1-30.0 & 17$^{\circ}$          & 48$^{\circ}$  \\
ESO 574 G33     & E$\gamma$       & SB(rs)bc    & 39.9$\pm$1.0 & 0.37$\pm$0.02  & 16.2-33.2; 15.3-34.1; 17.1-32.3 & 104$^{\circ}$         & 49$^{\circ}$  \\
ESO 576 G51     & E$\beta$        & SB(s)bc     & 30.4$\pm$1.9 & 0.47$\pm$0.04  & 18.0-39.1; 18.0-40.0; 18.0-38.2   & 80$^{\circ}$          & 46$^{\circ}$  \\
ESO 583 G2      & E$\beta$        & SB(rs)bc    & 28.4$\pm$1.0 & 0.47$\pm$0.04  & 10.0-27.8; 10.0-28.7; 10.0-26.9 & 52$^{\circ}$          & 44$^{\circ}$  \\
ESO 583 G7      & L$\beta$        & SB(rs)c     & 17.7$\pm$2.0 & 0.54$\pm$0.02  & 16.8-36.4; 15.9-37.3; 17.7-35.5 & 40$^{\circ}$          & 40$^{\circ}$  \\
ESO 602 G25     & L$\beta$        & SA(r)b      & 21.8$\pm$2.0 & 0.45$\pm$0.02  & 8.6-32.3; 7.7-33.2; 9.5-31.4    & 171$^{\circ}$         & 62$^{\circ}$  \\
ESO 606 G11     & H4$\beta$       & SB(rs)bc    & 25.2$\pm$2.4 & 0.50$\pm$0.03  & 10.9-21.8; 10.0-22.7; 11.8-20.9 & 85$^{\circ}$          & 46$^{\circ}$  \\
IC 1330         & E$\gamma$       & Sc? sp      & 37.8$\pm$1.1 & 0.35$\pm$0.02  & 18.0-27.3; 18.0-28.2; 18.0-26.4 & 60$^{\circ}$          & 59$^{\circ}$  \\
NGC 2584        & E$\beta$        & SB(s)bc?    & 29.7$\pm$3.0 & 0.50$\pm$0.04  & 12.7-28.2; 11.8-29.1; 13.6-27.3 & 17$^{\circ}$          & 57$^{\circ}$  \\
NGC 2722        & H4$\beta$       & SA(rs)bc    & 32.8$\pm$3.4 & 0.46$\pm$0.03  & 7.3-22.8; 6.4-23.7; 8.2-21.9    & 105$^{\circ}$         & 49$^{\circ}$  \\
NGC 3456        & E$\gamma$       & SBc         & 38.0$\pm$0.6 & 0.31$\pm$0.02  & 7.7-25.5; 6.8-26.4; 8.6-24.6    & 80$^{\circ}$          & 47$^{\circ}$  \\
NGC 7677        & E$\alpha$       & SAB(r)bc    & 17.0$\pm$0.8 & 0.66$\pm$0.02  & 20.0-37.8; 20.0-38.7; 20.0-36.9 & 35$^{\circ}$          & 51$^{\circ}$  \\
UGC 14          & L$\beta$        & Sc+         & 20.9$\pm$1.3 & 0.53$\pm$0.03  & 8.2-20.5; 7.3-21.4; 9.1-19.6    & 32$^{\circ}$          & 46$^{\circ}$  \\
UGC 210         & E$\beta$        & Sb          & 26.0$\pm$0.8 & 0.55$\pm$0.03  & 6.8-14.6; 5.9-15.5; 7.7-13.7    & 19$^{\circ}$          & 60$^{\circ}$  \\
\hline
\end{tabular}
\end{table*}

\section{Discussion}

We now discuss the results for each galaxy individually:

\noindent
{\em ESO 515-G3}: This galaxy is given a dust--penetrated class of L$\gamma$, 
due to its near-infrared pitch angle of 47.8$^{\circ}\pm$0.7 and its dominant $m=1$ Fourier 
mode. It shows narrow $m=1$ and $m=2$ components in its Fourier spectra, with 
the $m=2$ component also being very strong. The breadth of the Fourier 
components is indicative of the accuracy with which the observed spiral
structure can be approximated by a logarithmic spiral. For broader Fourier 
components, a logarithmic spiral is less applicable. This galaxy has a
rising rotation curve with a shear rate of 0.27$\pm$0.01. 

\noindent
{\em ESO 574-G33}: This galaxy is given a dust--penetrated class of 
E$\gamma$, due to its near-infrared pitch angle of 39.9$^{\circ}\pm$1.0 and its dominant 
$m=2$ Fourier mode. It shows narrow $m=1$ and $m=2$ components in its Fourier 
spectra, with the $m=1$ component also being very strong. It has a rising 
rotation curve with a shear rate of 0.37$\pm$0.02. 

\noindent
{\em ESO 576-G51}: This galaxy is given a dust--penetrated class of E$\beta$, 
due to its near-infrared pitch angle of 30.4$^{\circ}\pm$1.9 and its dominant $m=2$ Fourier 
mode. It shows narrow $m=1$ and $m=2$ components in its Fourier spectra, with 
the $m=1$ component also being very strong. Its rotation curve is approximately
flat with a shear rate of 0.47$\pm$0.04. 

\noindent
{\em ESO 583-G2}: This galaxy is given a dust--penetrated class of E$\beta$, 
due to its near-infrared pitch angle of 28.4$^{\circ}\pm$1.0 and its dominant $m=2$ Fourier 
mode. It shows a narrow $m=2$ component in its Fourier spectra. Its rotation
curve is approximately flat with a shear rate of 0.47$\pm$0.04.

\noindent
{\em ESO 583-G7}:  This galaxy is given a dust--penetrated class of 
L$\beta$, due to its near-infrared pitch angle of 17.7$^{\circ}\pm$2.0 and its 
dominant $m=1
$ Fourier mode. It shows a narrow $m=1$ component in its Fourier spectra. Its 
shear rate is 0.54$\pm$0.02 and its rotation curve remains approximately flat.

\noindent
{\em ESO 602-G25}: This galaxy is given a dust--penetrated class of E$\beta$, 
due to its near-infrared pitch angle of 21.8$^{\circ}\pm$2.0 and its dominant $m=2$ Fourier 
mode. It shows a narrow $m=2$ component in its Fourier spectra, and also a 
strong, but double--peaked $m=1$ component. Its rotation curve is approximately
flat with a shear rate of 0.45$\pm$0.02. 

\noindent
{\em ESO 606-G11}: This galaxy is given a dust--penetrated class of H4
$\beta$, due to its near-infrared pitch angle of 25.2$^{\circ}\pm$2.4 and its dominant $m=4
$ Fourier mode. It shows narrow $m=1$ and $m=4$ components and a double--peaked
$m=2$ component in its Fourier spectra, all of which appear as relatively 
strong modes. Its rotation curve is flat with a shear rate of 0.50$\pm$0.03.

\noindent
{\em IC 1330}: This galaxy is given a dust--penetrated class of E$\gamma$, 
due to its near-infrared pitch angle of 37.8$^{\circ}\pm$1.1 and its dominant $m=2$ Fourier 
mode. It shows a slightly broader than expected $m=2$ mode, which may be due 
to confusion with the strong bar in this galaxy. Its rotation curve is rising
with a shear rate of 0.35$\pm$0.02.

\noindent
{\em NGC 2584}: This galaxy is given a dust--penetrated class of E$\beta$, 
due to its near-infrared pitch angle of 29.7$^{\circ}\pm$3.0 and its dominant $m=2$ Fourier 
mode. It shows a narrow $m=2$ mode, which is more than twice as strong as any 
other mode. Its rotation curve is flat with a shear rate of 0.50$\pm$0.04.

\begin{figure}
\includegraphics{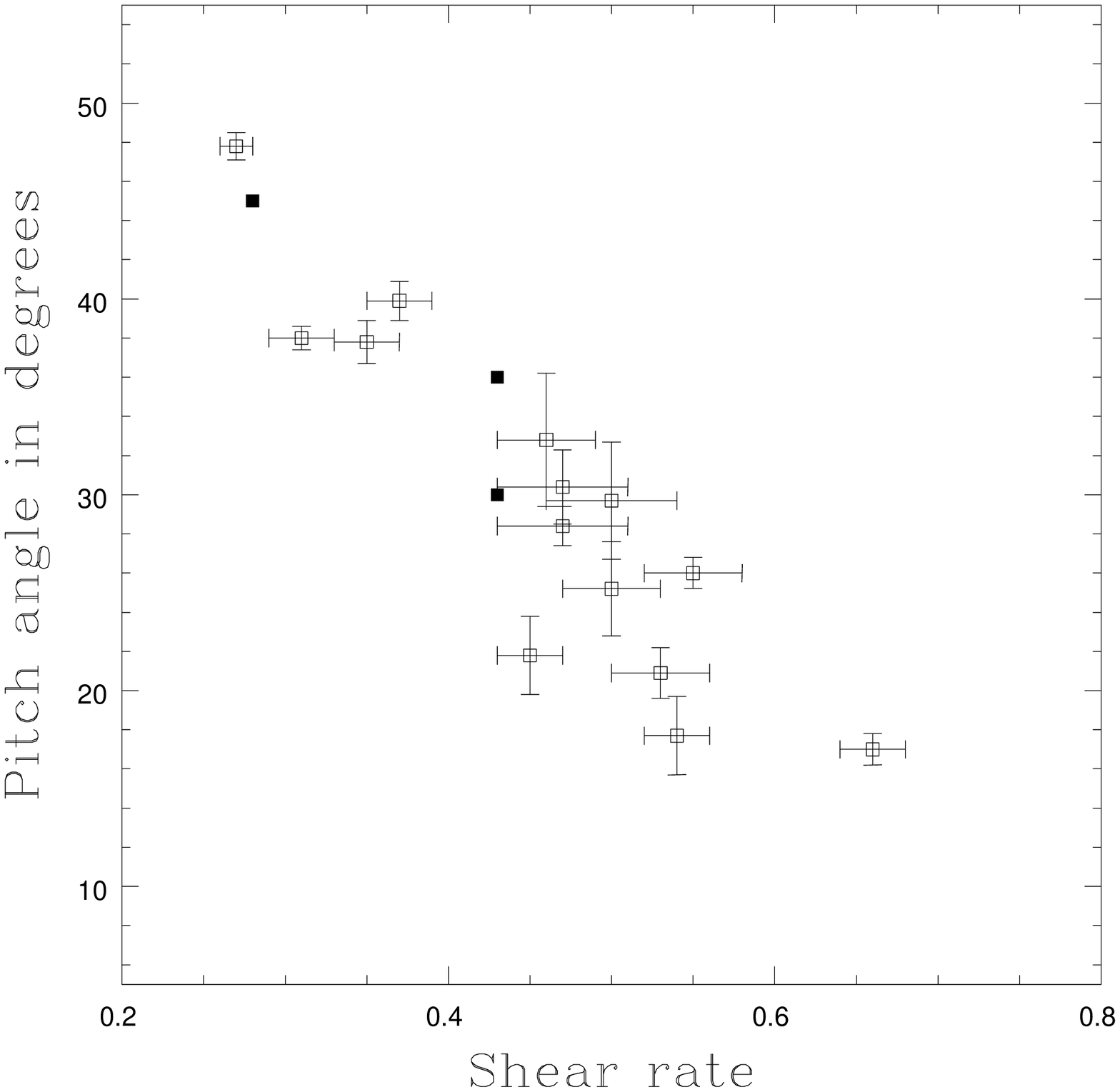}
\vspace*{9cm}
\caption{Near-infrared pitch angle in degrees versus shear rate. Hollow squares represent the
14 galaxies presented here. Solid squares are 3 galaxies presented in Block et al.
(1999).}
\end{figure}

\begin{figure}
\includegraphics{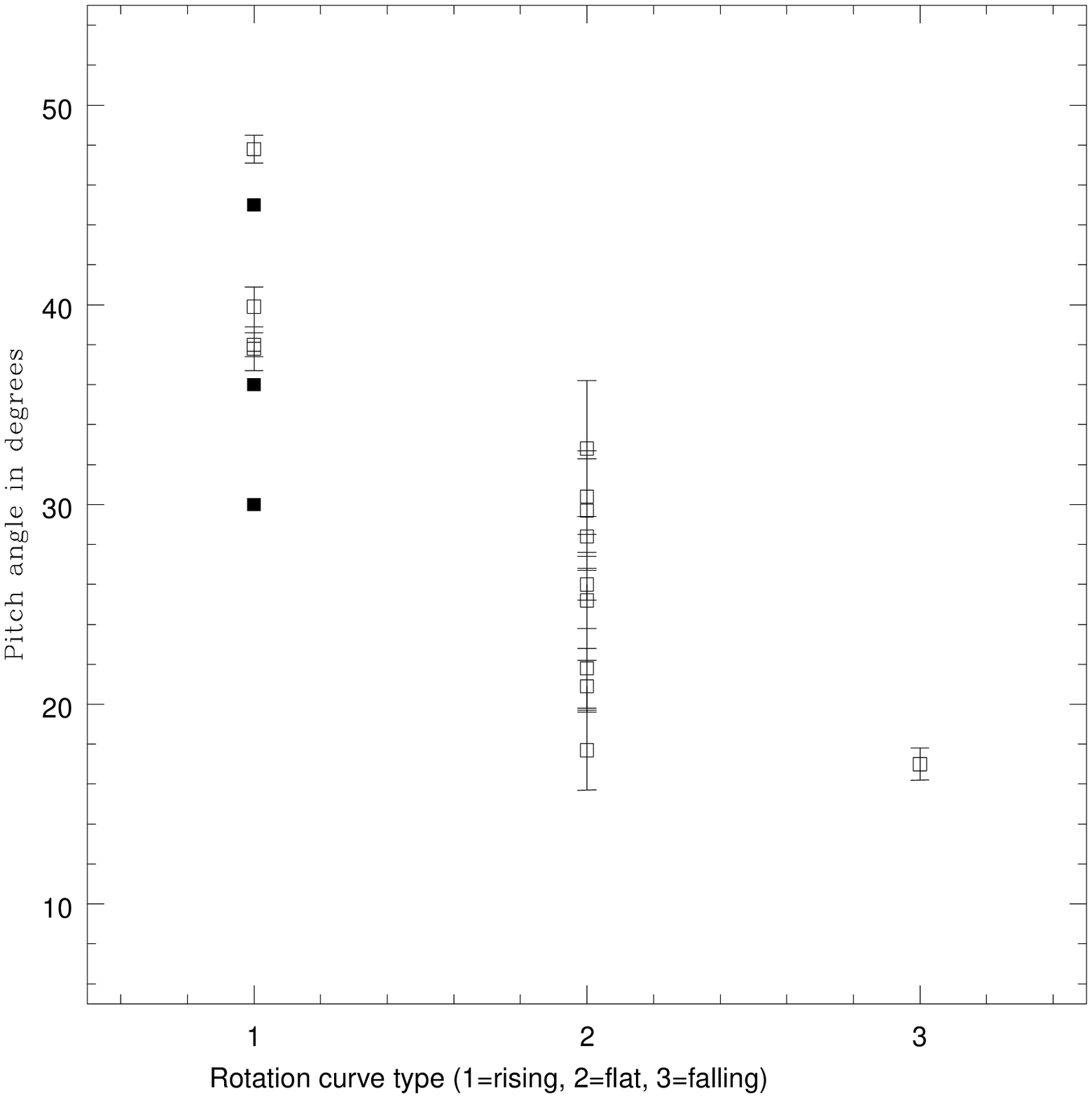}
\vspace*{9cm}
\caption{Near-infrared pitch angle in degrees versus rotation curve type. The symbols are 
the same as Figure 3}
\end{figure}

\noindent
{\em NGC 2722}: This galaxy is given a dust--penetrated class of H4$\beta$, 
due to its near-infrared pitch angle of 32.8$^{\circ}\pm$3.4 and its dominant $m=4$ Fourier 
mode. It shows a very broad modes from $m=1$ to $m=4$, which is probably due 
to the flocculent nature of its spiral structure, even when viewed in the 
near--infrared. Its rotation curve is approximately flat with a shear rate of 
0.50$\pm$0.03.

\noindent
{\em NGC 3456}: This galaxy is given a dust--penetrated class of E$\gamma$, 
due to its near-infrared pitch angle of 38.0$^{\circ}\pm$0.6 and its dominant $m=2$ Fourier 
mode. It shows a very narrow $m=2$ component, which is more than twice as 
strong as any other mode. Its rotation curve is rising with a shear rate of 
0.31$\pm$0.02.

\noindent
{\em NGC 7677}: This galaxy is given a dust--penetrated class of E$\alpha$, 
due to its near-infrared pitch angle of 17.0$^{\circ}\pm$0.8 and its dominant $m=2$ Fourier 
mode. It shows a very narrow $m=2$ component, which is more than three times 
as strong as any other mode. Its rotation curve is falling with a shear rate 
of 0.66$\pm$0.02.

\noindent
{\em UGC 14}: This galaxy is given a dust--penetrated class of L$\beta$, due 
to its near-infrared pitch angle of 20.9$^{\circ}\pm$1.3 and its dominant $m=1$ Fourier 
mode. It shows a  narrow $m=2$ component, with a double--peaked $m=2$ mode. 
The double-peaked $m=2$ mode suggests that this galaxy has both a leading
and a trailing $m=2$ component. From the spectra, these components have
very similar pitch angles, which explains why the error on the estimated
pitch angle is still quite low.
Its rotation curve has a shear rate of 0.53$\pm$0.03, due to its flat shape and
this suggests a spiral of intermediate winding
angle, as indicated by its dust--penetrated class. 

\noindent
{\em UGC 210}: This galaxy is given a dust--penetrated class of E$\beta$, due 
to its near-infrared pitch angle of 26.0$^{\circ}\pm$0.8 and its dominant $m=2$ Fourier 
mode. It shows a  narrow $m=2$ component, with a secondary, but weaker peak. 
Its $m=1$ component is also fairly strong. Its rotation curve remains 
approximately flat with a shear rate of 0.55$\pm$0.03. 

Figure 4 shows a plot of the shear rate versus the
near-infrared spiral arm pitch angle. As well as presenting a relatively tight correlation
(correlation coefficient = 0.93; significance = 99.9\%), it is also
interesting to note how spiral galaxies seem to fall into 3 distinct areas
on this plot, according to both their shear rates and their pitch
angles. Galaxies
with high shear rates (falling rotation curves)
and tightly wound spiral structure are found in the
bottom right and belong to the quantitative  $\alpha$ 
bin in the dust penetrated
class. 
Galaxies
with shear rates of approximately 0.5 (flat)
and intermediately wound spiral structure lie in the
middle and belong to the $\beta$ bin. Finally, the top left 
contains those galaxies with
loosely wound structure and low rates of shear (associated with
rising rotation curves).
These spirals belong to the $\gamma$ form family.

Figure 5 shows a plot between rotation
curve type (Burstein \& Rubin 1985), i.e. rising, flat or falling,
versus spiral arm pitch angle. A tight correlation (wherein the  
correlation coefficient is 0.86 and the significance 99.9\%) is found.

\section{Conclusions}

The shape of a rotation curve, beyond the turnover radius, 
is determined largely by the amount and
distribution of matter (including dark matter) contained in a 
spiral galaxy. 
The increase of shear rates from low to high, dictates 
mass distributions from small to large central mass concentrations. 
The correlation found
between pitch angle and shear rate is interpreted as follows:
galaxies
with higher rates of shear present a larger central mass concentration 
and more tightly wound arms. In contrast, open arm morphologies in the
dust penetrated, near-infrared regime are associated with rising
rotation curves, lower rates of galactic shear, and lower central mass
concentrations. 
This correlation was alluded to by the pioneering work of Lin \& Shu
(1964) and by the later spectroscopic studies of Burstein \& Rubin
(1986). It is in agreement with modal theories of spiral
structure (e.g. Bertin et al. 1989a, b; Bertin \& Lin 1996) and other 
numerical models based on the modal theory (e.g. Fuchs 1991, 2000). 

\section*{Acknowledgments}

The United Kingdom Infrared Telescope (UKIRT) is operated by the
Joint Astronomy Centre on behalf of the U.K. Particle Physics and
Astronomy Research Council (PPARC). DLB is indebted to Mrs M Keeton
and the Board of Trustees of the Anglo American Chairman's Fund for
their continued encouragement and support. This research utilized the
NASA/IPAC Extragalactic Database (NED), operated by the Jet
Propulsion Laboratory, California Institute of Technology, under contract
with the National Aeronautics and Space Administration. The authors 
wish to thank Paul Eskridge and George Lake for useful suggestions, and
the anonymous referee for comments which greatly improved the content
of this paper.

\end{document}